\documentclass[reprint,noshowpacs,noshowkeys,prd,balancelastpage,nofootinbib]{revtex4}
\usepackage{amsfonts}
\usepackage{mdframed}
\usepackage{amssymb}
\usepackage{footnote}
\usepackage{amsmath}
\usepackage{graphicx}
\usepackage{float}
\usepackage[font={footnotesize,it}]{caption}
\usepackage[utf8]{inputenc}
\usepackage{natbib}
\usepackage{tikz}
\usepackage{tikz-3dplot}
\usepackage[colorlinks=true,
            linkcolor=purple,
            urlcolor=purple,
            citecolor=blue]{hyperref}

\begin{document}

\title{Nonspherically-symmetric black hole in Einstein-massless scalar
theory }
\author{S. Habib Mazharimousavi}
\email{habib.mazhari@emu.edu.tr}
\affiliation{Department of Physics, Faculty of Arts and Sciences, Eastern Mediterranean
University, Famagusta, North Cyprus via Mersin 10, T\"{u}rkiye}
\date{\today }

\begin{abstract}
We introduce a two-parameter static, nonspherically-symmetric black hole
solution in the Einstein theory of gravity coupled with a massless scalar
field. The scalar field depends only on the polar coordinate $\theta$ in the
spherical coordinates representation. The scalar invariant of the spacetime,
namely, the Kretschmann scalar reveals that the black hole is singular on
its axis of symmetry for all nonzero values of the scalar charge. It also
manifests two surfaces of singularity for a certain interval of the value of
the scalar charge. These singularities are hidden behind the event horizon's
surface except the singularity on the axis which is naked. The energy-momentum
of the scalar field satisfies all energy conditions. A simple
investigation reveals that the circular orbit of massive and massless
particles in the equatorial plane is identical to the Schwarzschild black
hole.
\end{abstract}

\maketitle

\section{Introduction}

Fisher-Janis-Newman-Winicour (FJNW) metric is the exact solution of
Einstein's equations in the theory of gravity coupled minimally with a
massless scalar field. The solution is a one-parameter asymptotically flat,
static, and spherically symmetric with a naked singularity. It was first
introduced by Fisher in 1948 \cite{Fisher} and somehow was rediscovered by
Janis, Newman, and Winicour in 1968 \cite{JNW}. Furthermore, Wyman in 1981
introduced implicitly and independently the same spacetime \cite{Wyman} in
completing the early attempt of Bergmann and Leipnik \cite{BL} to obtain the
spherically symmetric solution of Einstein's equations with a special form
of the energy-momentum tensor. In \cite{Wyman} the uniqueness of the
solution was also proved. The solution introduced by Wyman in his original
paper doesn't contain the FJNW metric and it was shown by Virbhadra in 1997
that these solutions are the same \cite{VIR}. Their solution was also
reported in some other works \cite{J1,J2,J3,X,R}. This solution was
generalized by Xanthapoulos and Zannias to arbitrary spacetime dimensions in
1989 in \cite{X} where the fact that the spacetime is naked singular in 3+1
dimensions was generalized to any arbitrary dimensions. A deep analysis of
the d-dimensional FJNW spacetime was carried out by Abdolrahimi and Shoom in 
\cite{AS}. In 1993, Robert obtained the same solution in null coordinates
and showed that the asymptotic flatness of the static spherically-symmetric
solution of the Einstein equations in this theory is not a required
pre-assumption but it is intrinsic to the solution \cite{R}. In other words,
the FJNW solution is the most general static and spherically symmetric
solution of the theory. In \cite{J3}, FJNW was studied in the Newman-Penrose
formalism. The naked singularity of FJNW spacetime made it a good candidate
to examine its observational characteristic properties and compare them with
the correspondence black hole i.e., the Schwarzschild spacetime \cite%
{O1,O2,O3,O4,O5,O6,O7,O8,O9,O10,O11,O12,O13}. A good introduction to this
theory can be found in \cite{O14}. The gravitational collapse of a massless
scalar field was also studied in \cite{GC1,GC2,GC3,GC4,GC5}. Particularly,
we should mention the recent works on Einstein's gravity coupled with a
massless scalar field by Turimov, et al  \cite{A1,A2} as well as \cite{A3}
where in the latter one the electromagnetic field has also been considered.

Keeping in mind the impact of the FJNW metric in the evolution of our
understanding about the formation of singularities in general relativity and
their physical effects, we aim in this Letter to present a new solution in
the same context whose spherically symmetric feature is broken. In this
regard, we introduce a nonspherically-symmetric compact black hole solution
in Einstein's theory coupled with a nonspherically-symmetric massless scalar
field.

\section{The action and the solution}

\begin{figure}[tbp]
\includegraphics[width=100mm,scale=1]{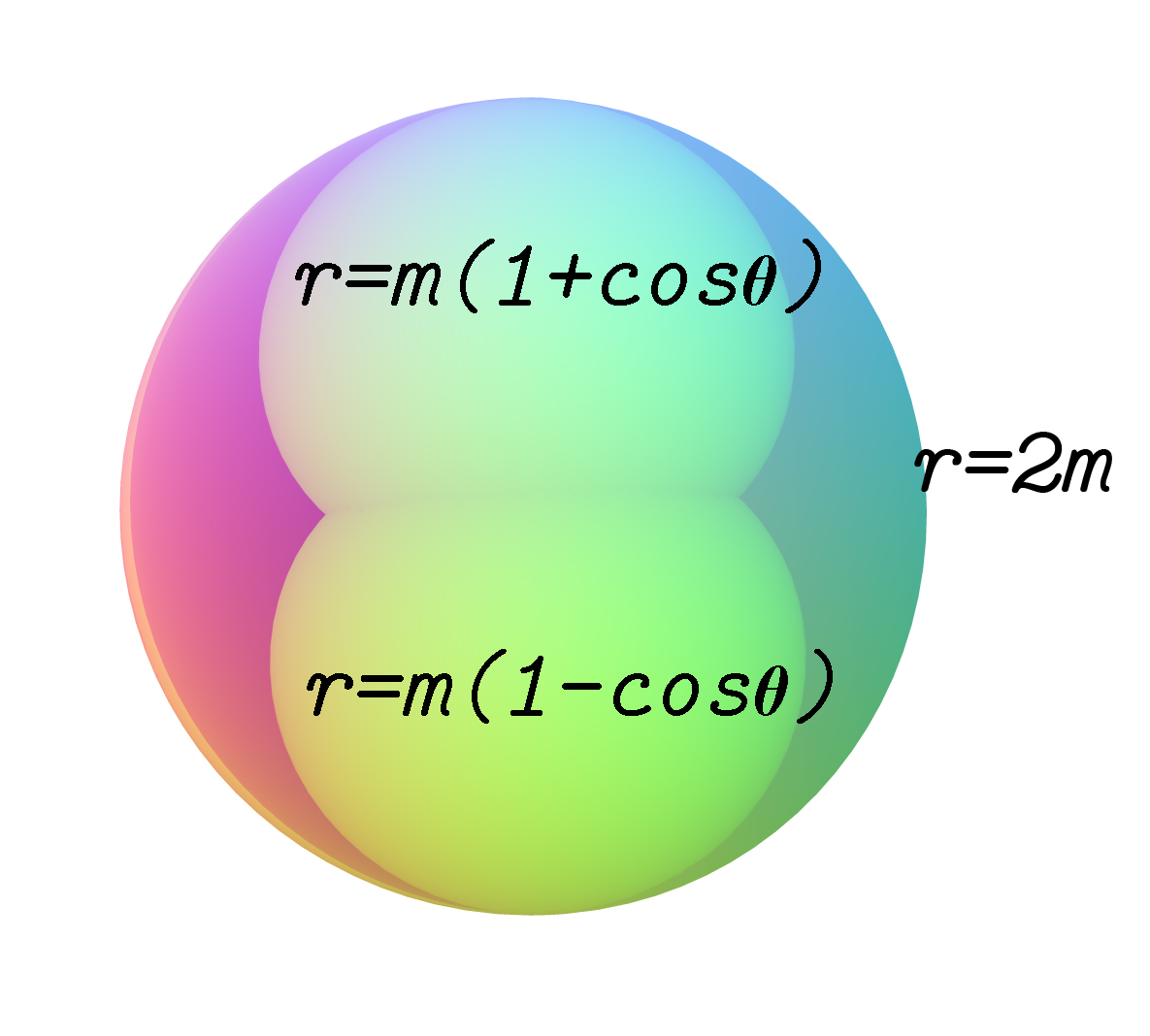}
\caption{The plots of the singular hypersurfaces for $4\protect\beta %
^{2}-1<0 $ which are the inner surfaces. The outer spherical surface is
defined by $r=2m$ which is nothing but the surface of the event horizon. The
singular hypersurfaces are hidden behind the event horizon. Furthermore, the
axis of symmetry is also a singular line.}
\label{F1}
\end{figure}

We start with the action of Einstein's theory coupled with a massless scalar
field ($8\pi G=1$) 
\begin{equation}
I=\frac{1}{2}\int d^{4}x\sqrt{-g}\left( \mathcal{R}-\left( \nabla \phi
\right) ^{2}\right) ,  \label{1}
\end{equation}%
where $\mathcal{R}$ is the Ricci scalar and $\phi $ is a massless scalar
field. Variation of the action w.r.t the metric tensor and the scalar field
yields, respectively, the Einstein and scalar field equations given by 
\begin{equation}
R_{\mu \nu }=\partial _{\mu }\phi \partial _{\nu }\phi ,  \label{2}
\end{equation}%
and%
\begin{equation}
\nabla _{\mu }\nabla ^{\mu }\phi =0.  \label{3}
\end{equation}%
Moreover, we consider the spacetime to be axially symmetric with the line
element 
\begin{equation}
ds^{2}=-f\left( r\right) dt^{2}+K\left( r,\theta \right) \left( \frac{dr^{2}%
}{f\left( r\right) }+r^{2}d\theta ^{2}\right) +r^{2}\sin ^{2}\theta d\varphi
^{2},  \label{4}
\end{equation}%
in which $f\left( r\right) $ and $K\left( r,\theta \right) $ are two metric
functions to be found. Herein, $\left\{ t\in \left( -\infty ,+\infty \right)
,r\in \left[ 0,\infty \right) ,\theta \in \left( 0,\pi \right) ,\varphi \in %
\left[ 0,2\pi \right) \right\} $ are the standard spherical coordinates such
that with $f\left( r\right) $ and $K\left( r,\theta \right) $ equal to
unity, (\ref{4}) reduces to the flat Minkowski spacetime in the spherical
coordinates representation. Concerning the line element (\ref{4}) the
nonzero components of the Ricci tensor are given by%
\begin{equation}
R_{tt}=\frac{f\left( rf^{\prime \prime }+2f^{\prime }\right) }{2K},
\label{5}
\end{equation}%
\begin{equation}
R_{rr}=-\frac{f^{\prime \prime }}{2f}+\frac{1}{2r^{2}f}\left( \left( \frac{%
K_{,\theta }}{K}\right) ^{2}-\frac{K_{,\theta \theta }}{K}-\frac{K_{,\theta }%
}{K\tan \theta }\right) +\frac{1}{2}\left( \left( \frac{K_{,r}}{K}\right)
^{2}-\frac{K_{,rr}}{K}\right) ,  \label{6}
\end{equation}%
\begin{equation}
R_{r\theta }=\frac{K_{,\theta }}{4K}\left( \frac{f^{\prime }}{f}+\frac{2}{r}%
\right) +\frac{K_{,r}}{2K\tan \theta },  \label{7}
\end{equation}%
\begin{equation}
R_{\theta \theta }=1-f-rf^{\prime }-\frac{K_{,r}}{2K}\left( r^{2}f^{\prime
}+2rf\right) +\frac{r^{2}f}{2}\left( \left( \frac{K_{,r}}{K}\right) ^{2}-%
\frac{K_{,rr}}{K}\right) +\frac{1}{2}\left( \frac{K_{,\theta }}{K}\right)
^{2}-\frac{K_{,\theta \theta }}{2K}+\frac{K_{,\theta }}{2K\tan \theta },
\label{8}
\end{equation}%
and%
\begin{equation}
R_{\varphi \varphi }=-\frac{\sin ^{2}\theta }{K}\left( rf^{\prime
}-1+f\right) ,  \label{9}
\end{equation}%
where $^{\prime }=\frac{d}{dr},$ $^{\prime \prime }=\frac{d^{2}}{dr^{2}}$, $%
K_{,r}=\frac{\partial K}{\partial r},K_{,\theta }=\frac{\partial K}{\partial
\theta },$ $K_{,rr}=\frac{\partial ^{2}K}{\partial r^{2}}$ and $K_{,\theta
\theta }=\frac{\partial ^{2}K}{\partial \theta ^{2}}.$ By assuming a
nonspherically-symmetric scalar field $\phi =\phi \left( \theta \right) $
i.e., the scalar potential depends only on the polar coordinate $\theta,$
one obtains exact solutions of the field equations expressed by%
\begin{equation}
f\left( r\right) =1-\frac{2m}{r},  \label{10}
\end{equation}%
\begin{equation}
K\left( r,\theta \right) =\frac{\left( \frac{m^{2}\sin ^{2}\theta }{r^{2}}%
\right) ^{2\beta ^{2}}}{\left( 1-\frac{2m}{r}+\frac{m^{2}\sin ^{2}\theta }{%
r^{2}}\right) ^{2\beta ^{2}}},  \label{11}
\end{equation}%
and%
\begin{equation}
\phi \left( \theta \right) =\pm \beta \ln \left( \frac{1-\cos \theta }{%
1+\cos \theta }\right) ,  \label{12}
\end{equation}%
in which $m$ and $\beta $ are two integration constants representing the
gravitational mass and the scalar field, respectively. Since the sign of $%
\pm \beta $ doesn't change the physical properties of the solution without
loss of generality we consider the positive branch and $\beta \geq 0.$ The
line element, hence, reads%
\begin{equation}
ds^{2}=-\left( 1-\frac{2m}{r}\right) dt^{2}+\frac{\left( \frac{m^{2}\sin
^{2}\theta }{r^{2}}\right) ^{2\beta ^{2}}}{\left( 1-\frac{2m}{r}+\frac{%
m^{2}\sin ^{2}\theta }{r^{2}}\right) ^{2\beta ^{2}}}\left( \frac{dr^{2}}{1-%
\frac{2m}{r}}+r^{2}d\theta ^{2}\right) +r^{2}\sin ^{2}\theta d\varphi ^{2}
\label{13}
\end{equation}%
which clearly is a 2-parameter, namely $m,$ and $\beta $, metric that
reduces to the standard Schwarzschild black hole in the absence of the
scalar field i.e., $\beta \rightarrow 0$. The Kretschmann scalar is
calculated to be 
\begin{multline}
\mathcal{K}=\frac{48\left( r^{2}-2mr+m^{2}\sin ^{2}\theta \right) ^{4\beta
^{2}-1}}{\left( \sin \theta \right) ^{4\left( 2\beta ^{2}+1\right)
}m^{8\beta ^{2}}r^{6}}\left( m^{4}\sin ^{6}\theta +\frac{2rm^{3}\left(
4\beta ^{2}-3\right) \sin ^{4}\theta }{3}+\right.  \label{14} \\
\left. \frac{r^{2}m^{2}\sin ^{2}\theta \left( \left( 6\beta ^{2}-3\right)
\cos ^{2}\theta +11\beta ^{4}-10\beta ^{2}+3\right) }{3}+\frac{2r^{3}m\beta
^{2}\left( \left( 2\beta ^{2}-1\right) \cos ^{2}\theta -5\beta ^{2}+1\right) 
}{3}+\beta ^{4}r^{4}\right) .  \notag
\end{multline}%
The latter expression reveals that for $4\beta ^{2}-1\geq 0$ or equivalently 
$\beta \geq \frac{1}{2},$ the Kretschmann scalar diverges at $\theta =0,\pi
, $ and $r=0$ which are located on the axis of symmetry, namely, the $z$%
-axis in the corresponding cylindrical coordinates. On the other hand, with $%
4\beta ^{2}-1<0$ or equivalently $\beta <\frac{1}{2}$, the spacetime is
singular not only at $\theta =0,\pi ,$ and $r=0$ but also at the surface of $%
r^{2}-2mr+m^{2}\sin ^{2}\theta =0$ which is plotted in Fig. \ref{F1}. 

Having the singularities located on the axis of symmetry and hidden by the
hypersurface $r=2m$, (\ref{13}) admits a spherical event horizon located at $%
r=2m$ which is practically the same as the event horizon of the black hole
in the absence of the massless scalar field. This is worth noting that%
\begin{equation}
\lim_{r\rightarrow 2m}\frac{\left( \frac{m^{2}\sin ^{2}\theta }{r^{2}}%
\right) ^{2\beta ^{2}}}{\left( 1-\frac{2m}{r}+\frac{m^{2}\sin ^{2}\theta }{%
r^{2}}\right) ^{2\beta ^{2}}}=1,  \label{14.5}
\end{equation}%
indicating that the area of the black hole is the same as the corresponding
Schwarzschild black hole.

Introducing the prolate spheroidal coordinates given by $X=\frac{r}{m}-1\in %
\left[ -1,\infty \right) ,$ $Y=\cos \theta \in \left[ -1,1\right) ,$ $%
\varphi =\varphi \in \left[ 0,2\pi \right) $, and $t=t\in \left( -\infty
,\infty \right) ,$ (\ref{13}) becomes%
\begin{equation}
ds^{2}=-\left( \frac{X-1}{X+1}\right) dt^{2}+m^{2}\left( \frac{X+1}{X-1}%
\right) \left\{ \left( X^{2}-1\right) \left( \frac{1-Y^{2}}{X^{2}-Y^{2}}%
\right) ^{2\beta ^{2}}\left( \frac{dX^{2}}{X^{2}-1}+\frac{dY^{2}}{1-Y^{2}}%
\right) +\left( X^{2}-1\right) \left( 1-Y^{2}\right) d\varphi ^{2}\right\} .
\end{equation}%
Calculating the Kretschmann scalar in this coordinates system reveals%
\begin{equation}
\mathcal{K}=\frac{48}{m^{4}}\left( \frac{X^{2}-Y^{2}}{1-Y^{2}}\right)
^{4\beta ^{2}}\left( \frac{\beta ^{4}\left( X^{2}+\frac{4}{3}\left(
XY^{2}-X+1\right) -\frac{7}{3}Y^{2}\right) }{\left( X^{2}-Y^{2}\right)
\left( 1-Y^{2}\right) ^{2}\left( 1+X\right) ^{4}}+\frac{\frac{2}{3}\beta
^{2}\left( X^{2}-Y^{2}+3XY^{2}-3X\right) }{\left( X^{2}-Y^{2}\right) \left(
1-Y^{2}\right) \left( 1+X\right) ^{5}}+\frac{1}{\left( 1+X\right) ^{6}}%
\right)
\end{equation}%
which shows the singularities at $X=-1$ corresponding to $r=0$, $Y^{2}=1$
corresponding the poles, $X^{2}=Y^{2}$ corresponding to $\left( \frac{r}{m}%
-1\right) ^{2}=\cos ^{2}\theta $ provided $4\beta ^{2}<1.$ The scalar
potential is also transformed to be%
\begin{equation}
\phi \left( Y\right) =\beta \ln \left( \frac{1-Y}{1+Y}\right) ,  \label{SF}
\end{equation}%
which is clearly singular at $Y=\pm 1$ which is the boundary of the domain
of $Y.$

\section{The energy conditions}

The black hole introduced in the previous section is supported by a
nonspherically symmetric scalar field. In this section, we investigate the
energy conditions, namely the null, weak, strong, and dominant energy
conditions, abbreviated by NEC, WEC, SEC, and DEC, respectively. To do so we
calculate the energy-momentum of the scalar field given by%
\begin{equation}
T_{\mu }^{\nu }=\partial _{\mu }\phi \partial ^{\nu }\phi -\frac{1}{2}\delta
_{\mu }^{\nu }\partial _{\alpha }\phi \partial ^{\alpha }\phi
\end{equation}%
which knowing that $\phi \left( Y\right) $, (\ref{SF}) explicitly reads%
\begin{equation}
T_{\mu }^{\nu }=-\frac{1}{2}\left( \partial _{Y}\phi \partial ^{Y}\phi
\right) diag\left[ 1,1,-1,1\right]
\end{equation}%
where%
\begin{equation}
\partial _{Y}\phi \partial ^{Y}\phi =g^{YY}\left( \partial _{Y}\phi \right)
^{2}=\frac{4\beta ^{2}}{m^{2}}\frac{1}{\left( X+1\right) ^{2}}\frac{\left(
X^{2}-Y^{2}\right) ^{2\beta ^{2}}}{\left( 1-Y^{2}\right) ^{2\beta ^{2}+1}}.
\end{equation}%
Hence, introducing $T_{\mu }^{\nu }=diag\left[ -\rho ,p_{X},p_{Y},p_{\varphi
}\right],$ we implicitly find the energy density and the pressures in
different directions to be given by%
\begin{equation}
\rho =-p_{X}=p_{Y}=-p_{\varphi }=\frac{2\beta ^{2}}{m^{2}}\frac{1}{\left(
X+1\right) ^{2}}\frac{\left( X^{2}-Y^{2}\right) ^{2\beta ^{2}}}{\left(
1-Y^{2}\right) ^{2\beta ^{2}+1}}.  \label{EMT}
\end{equation}%
The NEC implies $\rho +p_{i}\geq 0$ which clearly satisfied. Furthermore,
the WEC is the union of NEC and $\rho \geq 0$ which is also perfectly
satisfied. Next is the SEC which is the union of NEC and $\rho
+p_{X}+p_{Y}+p_{\varphi }\geq 0$ which is certainly satisfied. Finally the
DEC implies $\rho \geq 0,$ and $\rho -\left\vert p_{i}\right\vert \geq 0$ in
which $i=X,Y,\varphi .$ With (\ref{EMT}) one can easily see that DEC is also
satisfied. Therefore, all energy conditions are perfectly satisfied by the
energy-momentum tensor of the scalar field that indicates the black hole is
supported by a physical scalar field. This is also important to mention that
the energy-momentum tensor is singular at $X=-1$ and $Y=\pm 1.$ The former
singular point is the center of the black hole corresponding to $r=0$ in the
spherical coordinates system. The latter implies the north and the south
poles which effectively addresses the axis of symmetry. On the surface of
singularity i.e., $X^{2}=Y^{2}$, the energy-momentum tensor vanishes which
indicates the nature of this singularity is not the energy-momentum tensor
or the scalar field but it is due to the geometry of the spacetime. It is in
analogy with the $\gamma $ metric \cite{ZV1,ZV2}.

\section{The photon orbit}

In this section, we investigate the effect of the scalar field on the photon
orbit around the black hole. The Lagrangian of a null particle moving in the
vicinity of the black hole (\ref{13}) is given by%
\begin{equation}
2\mathcal{L}=-\left( 1-\frac{2m}{r}\right) \dot{t}^{2}+\frac{\left( \frac{%
m^{2}\sin ^{2}\theta }{r^{2}}\right) ^{2\beta ^{2}}}{\left( 1-\frac{2m}{r}+%
\frac{m^{2}\sin ^{2}\theta }{r^{2}}\right) ^{2\beta ^{2}}}\left( \frac{\dot{r%
}^{2}}{1-\frac{2m}{r}}+r^{2}\dot{\theta}^{2}\right) +r^{2}\sin ^{2}\theta 
\dot{\varphi}^{2},  \label{15}
\end{equation}%
where a dot stands for the derivative with respect to an affine parameter.
The energy of the particle as well as its angular momentum about the axis of
symmetry are conserved such that%
\begin{equation}
E=-\frac{\partial \mathcal{L}}{\partial \dot{t}}=1-\frac{2m}{r},  \label{16}
\end{equation}%
and%
\begin{equation}
\ell =\frac{\partial \mathcal{L}}{\partial \dot{\varphi}}=r^{2}\sin
^{2}\theta \dot{\varphi}.  \label{17}
\end{equation}%
The equation of motion in $\theta $ direction is rather far from being
trivially satisfied simply by $\theta =\theta _{0}$ unless $\theta _{0}=%
\frac{\pi }{2}.$ Applying the constraint of the null geodesics i.e., $\dot{x}%
_{\mu }\dot{x}^{\mu }=0,$ on the equatorial plane where $\theta _{0}=\frac{%
\pi }{2}$ the main radial equation is obtained to be%
\begin{equation}
\left( \frac{r}{m}-1\right) ^{-4\beta ^{2}}\dot{r}^{2}+V_{eff}\left(
r\right) =E^{2},  \label{18}
\end{equation}%
where the effective potential is given by%
\begin{equation}
V_{eff}\left( r\right) =\frac{\ell ^{2}}{r^{2}}\left( 1-\frac{2m}{r}\right) .
\label{19}
\end{equation}%
We observe from the effective potential that the circular orbit where $\dot{r%
}=0$ and $\ddot{r}=0$ is possible at $r_{c}=3m$ where $V_{eff}^{\prime
}\left( r_{c}\right) =0$ however since $V^{\prime \prime }\left(
r_{c}\right) <0$ this circular orbit is unstable. The situation is the same
as the Schwarzschild black hole. Although the circular orbit of the null
particle is $\beta $-independent, the actual general motion of a null as
well as a massive particle very much depends on $\beta $ which is clearly
observed from (\ref{18}).

\section{Conclusion}

In this Letter, we introduced an exact black hole solution in the theory of
gravity coupled with a massless scalar field. The solution contains two
integration parameters which are $m$ and $\beta .$ These parameters are
related to the mass of the black hole and the charge of the scalar field. In
particular, $\beta \geq 0$ may be called the charge of the scalar field and
it plays a crucial role in the global configuration of the solution. The
contribution of $\beta $ is in analogy with the parameter $\gamma $ in $%
\gamma $-metric which is also known as Zipoy-Voorhees metric \cite{ZV1,ZV2}.
For $\beta =0$ the solution becomes the Schwarzschild black hole with a
timelike singularity at $r=0.$ For $0<\beta $ the solution is axially
symmetric and geometrically the spacetime is deformed such that $\beta =%
\frac{1}{2}$ seems to play a topological phase transition point such that
for $\beta \geq \frac{1}{2}$ the surfaces of singularities disappear. Hence,
topologically one defines three regions: i. $\beta =0$, ii. $0<\beta <\frac{1%
}{2}$ and ii. $\frac{1}{2}\leq \beta $. In all three cases, the event
horizon is formed at $r=2m$ although for $\beta >0$ the horizon surface is
singular at the poles. The scalar field depends only on the polar angle $%
\theta $ and is singular at the poles. Unlike, the well-known FJNW metric
which is spherically symmetric, asymptotically flat, and naked singular, the
solution we presented here is non-spherical, non-asymptotically flat with an
event horizon. The main singularity is the axis of symmetry of the black
hole in analogy with the Levi-Civita spacetime \cite{LC}, however, the event
horizon is spherical and covers the singularity. Except for the poles,
which are singular, the event horizon in other directions is regular. We
have also shown explicitly that the energy conditions are all satisfied
unconditionally. The photon orbit on the equatorial plane is exactly the
same as the Schwarzschild black hole.

\end{document}